\documentclass[pra,twocolumn,floatfix,longbibliography]{revtex4-1}

\usepackage{amssymb,amsmath,amstext}
\usepackage{graphicx}
\usepackage{epstopdf}
\usepackage{color}
\usepackage{bm}
\usepackage{appendix}
\usepackage[T1]{fontenc}
\usepackage{bbold}
\usepackage{bbm}
\usepackage{latexsym}
\usepackage[colorlinks=true,citecolor=blue,linkcolor=magenta]{hyperref}

\long\def\ca#1\cb{} 

\newcommand{\avg}[1]{\langle #1\rangle }
\newcommand{\ket}[1]{|#1\rangle}               
\newcommand{\bra}[1]{\langle #1|}              
\newcommand{\dya}[1]{\ket{#1}\!\bra{#1}}

\newcommand{\Tr}{{\rm Tr}}

\newcommand{\ave}[1]{\langle #1\rangle}               

\newcommand*{\id}{\openone}

\begin{document}
\title{Qubit-efficient exponential suppression of errors}
\author{Piotr Czarnik}
\affiliation{Theoretical Division, Los Alamos National Laboratory, Los Alamos, NM 87545, USA}
\author{Andrew Arrasmith}
\affiliation{Theoretical Division, Los Alamos National Laboratory, Los Alamos, NM 87545, USA}
\affiliation{Quantum Science Center, Oak Ridge, TN 37931, USA}
\author{Lukasz Cincio}
\affiliation{Theoretical Division, Los Alamos National Laboratory, Los Alamos, NM 87545, USA}
\affiliation{Quantum Science Center, Oak Ridge, TN 37931, USA}
\author{Patrick J. Coles}
\affiliation{Theoretical Division, Los Alamos National Laboratory, Los Alamos, NM 87545, USA}
\affiliation{Quantum Science Center, Oak Ridge, TN 37931, USA}
\begin{abstract}
Achieving a practical advantage with near-term quantum computers hinges on having effective methods to suppress errors. Recent breakthroughs have introduced methods capable of exponentially suppressing errors by preparing multiple noisy copies of a state and virtually distilling a more purified version. Here we present an alternative method, the Resource-Efficient Quantum Error Suppression Technique (REQUEST), that adapts this breakthrough to much fewer qubits by making use of active qubit resets, a feature now available on commercial platforms. Our approach exploits a space/time trade-off to achieve a similar error reduction using only $2N+1$ qubits as opposed to $MN+1$ qubits, for $M$ copies of an $N$ qubit state. Additionally, we propose a method using near-Clifford circuits to find the optimal number of these copies in the presence of realistic noise, which limits this error suppression. We perform a numerical comparison between the original method and our qubit-efficient version with a realistic trapped-ion noise model. We  find that REQUEST can reproduce the exponential suppression of errors of the virtual distillation approach, while out-performing virtual distillation when fewer than $3N+1$ qubits are available. Finally, we examine the scaling of the number of shots $N_S$ required for REQUEST to achieve useful corrections. We find that $N_S$ remains reasonable well into the quantum advantage regime where $N$ is hundreds of qubits.
\end{abstract}
\maketitle

\section{Introduction}

One of the most serious challenges in demonstrating a practical advantage for quantum computing over classical computing in the near term is the hardware noise~\cite{wang2020noise,franca2020limitations}. While many expect that fault-tolerant quantum computing will eventually become available, this possibility remains distant. Instead, we are approaching the arrival of noisy, intermediate scale quantum (NISQ) devices that employ hundreds or more noisy qubits~\cite{preskill2018quantum}.

A number of error mitigation methods have been proposed for this NISQ era. Perhaps the most prominent method is zero noise extrapolation, whereby the noise is increased in a controlled manner, allowing one to perform a function fit and then extrapolate to the noiseless limit~\cite{temme2017error,kandala2018error,dumitrescu2018cloud,endo2018practical,otten2019recovering, giurgica2020digital,he2020zero}. Alternatively, if one knows the noise model of the device being used, gates can be introduced probabilistically in order to cancel the effect of the noise on average~\cite{temme2017error}. A third approach is to leverage classically simulable quantum circuits to infer properties of the hardware noise by comparing classically simulated outputs with noisy results. With this approach, one can use regression to estimate the noiseless result of interest directly~\cite{czarnik2020error}, inform the functional form for a zero noise extrapolation~\cite{lowe2020unified}, or estimate the gate distributions for probabilistic error correction~\cite{strikis2020learning}. Circuit optimization by optimal compiling provides another error mitigation method by producing noise-resilient quantum circuits~\cite{cincio2018learning,cincio2020machine,murali2019noise,khatri2019quantum,sharma2019noise,cerezo2020variational}. Finally, a number of application specific approaches have been proposed, leveraging symmetries and/or post-selection techniques to mitigate errors~\cite{mcardle2019error,bonet2018low,otten2019noise,cai2021quantum}.

\begin{figure}[ht!]
\includegraphics[width=.70\columnwidth]{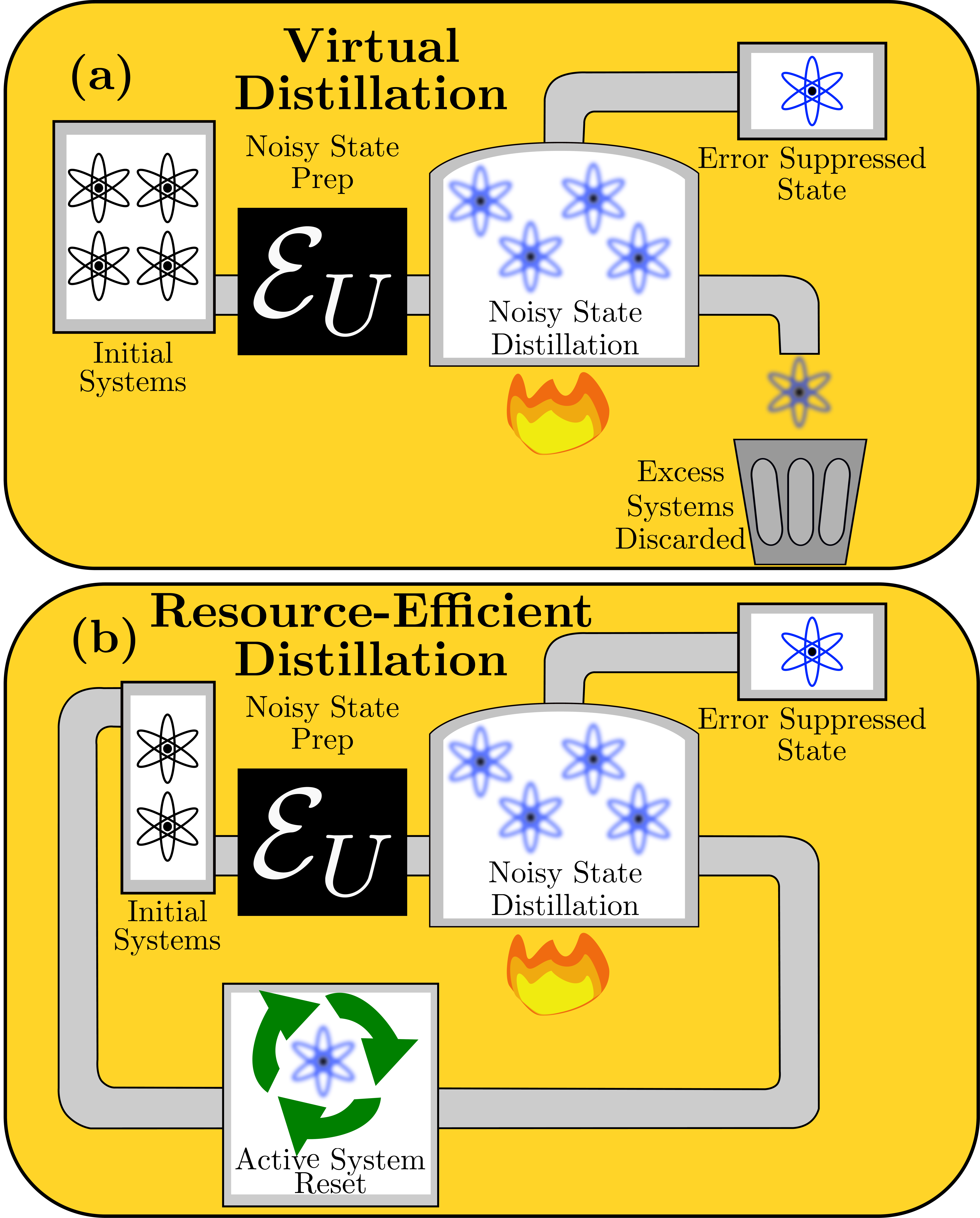}
\caption{\textbf{Schematic comparison of the  virtual distillation method to our resource-efficient version.} Virtual distillation, shown in \textbf{(a)}, takes a number of systems in a reference state and subjects them to a noisy state preparation described by the quantum channel $\mathcal{E}_U$. From these states, a single distilled copy is output while the other systems are discarded. The distilled copy has increased purity with respect to the noisy states which is improving exponentially with increasing number of the copies. The resource efficient version of this technique, shown in \textbf{(b)}, accomplishes the same process but uses active reset operations to recycle systems and thus only requires two systems to simulate the distillation with many systems.}
\label{fig:distill}
\end{figure}

In a recent breakthrough, a new approach was proposed that uses additional copies of a quantum state of interest to suppress errors~\cite{koczor2020exponential,huggins2020virtual}. These methods make use of the fact that, for a given density matrix $\rho$, the state $\rho^M/\Tr[\rho^M]$ approaches a pure state exponentially quickly with increasing $M$. Measurements on this state can be made by preparing $M$ copies of $\rho$, and this protocol has been termed the exponential suppression of quantum errors when $\rho$ represents a noisy state that results from attempting to prepare a given pure state~\cite{koczor2020exponential}. 

Nevertheless, the ability of this approach to exponentially suppress errors is subject to limitations. First, in general the pure state this method approaches may not be exactly the intended pure state, which provides a noise floor for the method. Additionally, the protocol includes the action of  noisy controlled swaps  on the copies. Therefore when one accounts for the increased number of  the controlled swaps associated with  larger $M$, there is a point where adding copies may increase the impact of noise more than it suppresses it~\cite{huggins2020virtual}.

A different limitation of this error suppression strategy comes from the number of qubits required. If the state $\rho$ requires $N$ qubits to prepare, then $MN+1$ qubits are required in total. For NISQ devices capable of achieving a quantum advantage, the required number of qubits may prove more limiting to the number of copies that can be used than the noise of the controlled swaps. 

We present an alternative approach below that uses the same framework to achieve an exponential error suppression (in the same sense) with only $2N+1$ qubits. To that end we use active qubit resets enabling reusing qubits during a quantum algorithm execution by re-initializing them in a known state \cite{reed2010fast,geerlings2013demonstrating,mcclure2016rapid,magnard2018fast}.  In that way, inspired by  qubit-efficient algorithms for Renyi entropy computation~\cite{yirka2020qubit},  we replace a circuit implementing the exponential suppression  by an equivalent one with  increased depth proportional to $M$ and  fixed width.  We call this new method the resource-efficient quantum error suppression technique (REQUEST).  We schematically compare REQUEST with the original exponential suppression methods in Fig.~\ref{fig:distill}.     As the qubit resets are enabled by major quantum computing architectures, including superconducting qubit~\cite{corcoles2021exploiting} and trapped-ion devices~\cite{foss2020holographic},  we expect the method to have a wide range of applications.

Below we first briefly review the error suppression formalism presented in Refs.~\cite{koczor2020exponential,huggins2020virtual}. Next, we introduce our modification, REQUEST, and present a method to estimate the optimal number of copies $M_{\textrm{opt}}$ using near-Clifford circuits. Additionally, we analyze the scaling of the number of state preparations and measurements required by REQUEST, and we find that this requirement remains reasonable even when $N$ is hundreds of qubits. We then present a numerical comparison of the performance of our method and the original. Finally, we present our conclusions and discuss future directions. 


\section{Exponential suppression of errors}

Recently, methods for the suppression of quantum errors by virtual distillation (VD) using $M$ copies of a state have been proposed~\cite{koczor2020exponential,huggins2020virtual}. These methods are based on the assumption that the desired pure state at the end of a unitary evolution will be close to the eigenvector of the density matrix with the largest eigenvalue. 
To make this precise, suppose that we have an initial state $\ket{\phi}$ on $N$ qubits that we wish to act on with a unitary $U$, but the device we are working with is noisy. Denoting by $\mathcal{E}_U$ the quantum channel that results from attempting to apply $U$, we end up preparing the state
\begin{equation}
\begin{aligned}
    \mathcal{E}_U(\ket{\phi}\bra{\phi})=&\rho \\
    =&\sum_{i=1}^{D} p_i \ket{\psi_i}\bra{\psi_i}.
\end{aligned}
\end{equation}
Here the eigenvalues $p_i$ are ordered in descending order (for convenience) and $D=2^N$ is the dimension of the Hilbert space. VD then works from the assumption that
\begin{equation}\label{eq:approximation}
    U\ket{\phi}\approx \ket{\psi_1}.
\end{equation}
 Note that if all errors introduced are orthogonal to the state of interest, the approximation in Eq.~\eqref{eq:approximation} becomes exact~\cite{huggins2020virtual}. 

With the approximation in Eq.~\eqref{eq:approximation} in mind, VD considers the quantity~\cite{koczor2020exponential,huggins2020virtual}
\begin{equation}\label{eq:trace_ratio}
\begin{aligned}
    \frac{\Tr[X\rho^M]}{\Tr[\rho^M]}=&\frac{\bra{\psi_1}X\ket{\psi_1}}{1+\sum_{i=2}^D (p_i/p_1)^M} \\
    &+\frac{\sum_{i=2}^D (p_i/p_1)^M\bra{\psi_i}X\ket{\psi_i}}{{1+\sum_{i=2}^D (p_i/p_1)^M}}\\
    =&\avg{X}_{\textrm{mitigated}}.
\end{aligned}
\end{equation}
So long as $p_1$ is larger than any other eigenvalue, this ratio approaches $\bra{\psi_1}X\ket{\psi_1}$ exponentially with $M$ and so is used to calculate the error suppressed expectation value of $X$, which we denote  $\avg{X}_{\textrm{mitigated}}$.

\begin{figure}[t]
\includegraphics[width=\columnwidth]{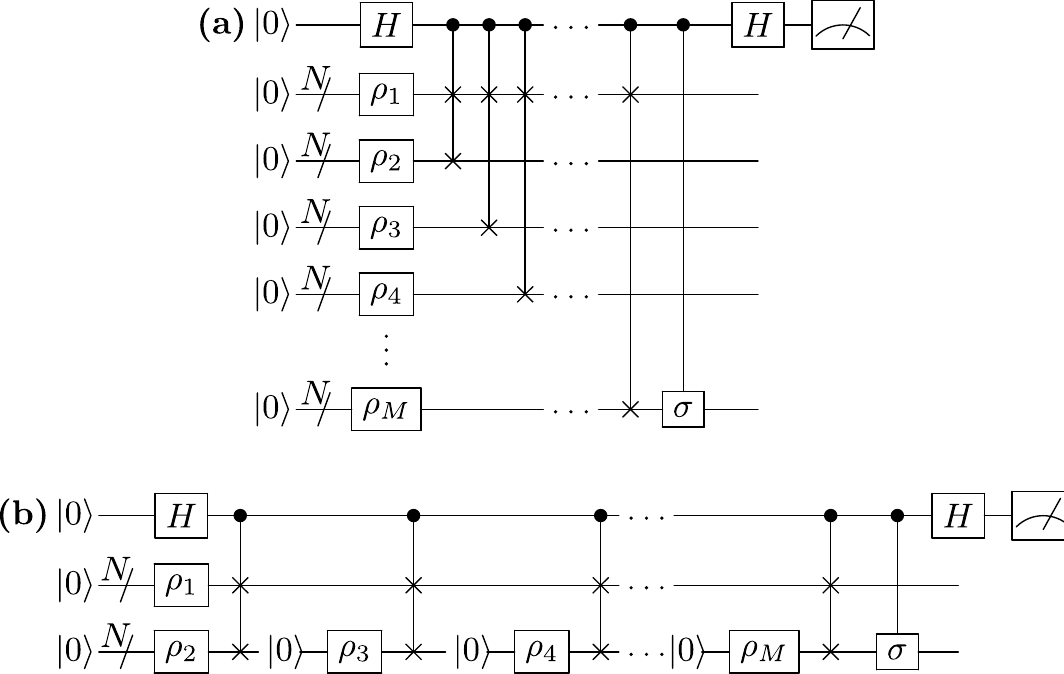}
\caption{\textbf{Circuit diagrams for the exponential suppression of errors with $M$ copies.} Here each $\rho_i$ denotes the circuit to prepare the $i^\mathrm{th}$ copy of the state $\rho$. Also, $\sigma=I$ or $\sigma=X$, where $X$ is the observable whose expectation value will be mitigated. Diagram \textbf{(a)} shows a circuit diagram like the one proposed by~\cite{koczor2020exponential} to suppress errors. Diagram \textbf{(b)} shows our alternative formulation using active qubit resets, which are represented by a break in a wire followed by $\ket{0}$.  
} 
\label{fig:circuit_diagrams}
\end{figure}

VD computes the numerator and denominator of Eq.~\eqref{eq:trace_ratio} by preparing $M$ copies of $\rho$ and one ancilla qubit, using a circuit like the one shown in Fig.~\ref{fig:circuit_diagrams}\textbf{(a)}. The main idea is to apply a controlled  derangement operation  commonly used in computing Renyi entropies~\cite{johri2017entanglement,subacsi2019entanglement}  to the copies (the derangement  is a permutation of copies which changes position of each copy). The derangement is implemented with the controlled swap gates.   
To find $\avg{X}_{\textrm{mitigated}}$
we apply a controlled $\sigma$ gate to the permuted copies  (see Fig.~\ref{fig:circuit_diagrams}\textbf{(a)}) and measure the ancilla  qubit. 
Denoting the probability of getting $0$ as the result of the measurement with $\sigma=X$  as $\textrm{prob}_0$ and with $\sigma =I$ (the identity) as $\textrm{prob}'_0$, we then have:
\begin{equation}\label{exp}
\avg{X}_{\textrm{mitigated}} =\frac{\Tr[X\rho^M]}{\Tr[\rho^M]}
=\frac{2\textrm{prob}_0-1}{2\textrm{prob}'_0-1}.
\end{equation}

Even on a quantum device with many qubits there are practical limitations to the error suppression offered by VD. First, many physical error channels will produce errors which are not orthogonal to $U\ket{\phi}$, worsening the approximation in Eq.~\eqref{eq:approximation}. This effect introduces a floor below which the error cannot be suppressed~\cite{huggins2020virtual}:
\begin{equation}\label{eq:floor}
    \epsilon=|\bra{\psi_1}X\ket{\psi_1}-\bra{\phi}U^\dagger X U \ket{\phi}| \ .
\end{equation}

Additionally, since the application of the controlled derangement is subject to error channels, in practical applications there will usually be a finite optimal number of copies ($M_{\textrm{opt}}$) that can be used before the additional error introduced outweighs the suppression. (We note that $M_{\textrm{opt}}$ may not be finite for highly idealized noise models such as global depolarizing noise, but it will be for realistic noise models based on current hardware.) As determining $M_{\textrm{opt}}$ would require detailed knowledge of the noise channels, this value can be difficult to predict in practice. Finally, we note that VD is robust to the copies of $\rho$ being imperfect (perhaps due to differences in their noise channels) if they still have the same $\ket{\psi_1}$ corresponding to the largest eigenvalue~\cite{koczor2020exponential}.

\section{Qubit-efficient error suppression}

In addition to the accumulation of hardware errors, the number of qubits available (as $N_{\textrm{tot}}=MN+1$ qubits are required) and the difficulty of entangling them limits the mitigation attainable with VD. Particularly for NISQ devices, the severely limited number of qubits and connectivity may well be more restrictive. 
Inspired by qubit-efficient methods for computing Renyi entropies~\cite{yirka2020qubit}, we therefore propose a variant VD we call the Resource-efficient Quantum Error Suppression Technique (REQUEST). REQUEST utilizes active qubit resets to reduce $N_{\textrm{tot}}$ to $2N+1$, independent of $M$. 

The prototypical circuit diagram for REQUEST is schematically depicted in Fig.~\ref{fig:circuit_diagrams}\textbf{(b)}. We note that the circuit diagrams in Fig.~\ref{fig:circuit_diagrams} are mathematically equivalent, though the noise channels that result from implementing them will differ. REQUEST therefore reduces $N_{\textrm{tot}}$ at the cost of increased circuit depth. This trade-off means that the idling time for the control qubit and one copy of $\rho$ is greatly increased in REQUEST as compared with VD. However, on devices with limited connectivity, the cost of performing the derangement operation on $M$ copies may offset this difficulty. In such case, REQUEST may prove a more efficient alternative, even if many qubits are available.

\begin{figure}[t]
\includegraphics[width=\columnwidth]{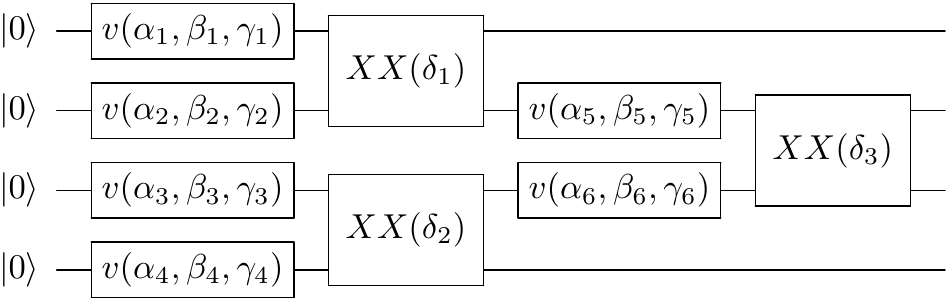}
\caption{\textbf{Random quantum circuit structure for our numerical benchmarks.}  Here we show an exemplary  random quantum circuit created with  one layer of a trapped-ion hardware efficient ansatz. The layer is built from two layers of alternating nearest-neighbor $XX(\delta)=e^{-i\delta \sigma_X^j \sigma_X^{j+1}}$ gates, where $\sigma_X^j$ is a Pauli operator acting on qubit~$j$. The $XX$ gates are decorated with general single-qubit unitaries  $v(\alpha,\beta,\gamma) = R_Z(\alpha) R_Y(\beta) R_Z(\gamma)$. Here $R_Z(\alpha) = e^{-i\alpha/2 \sigma_Z}$, $R_Y(\beta) = e^{-i\beta/2 \sigma_Y}$, and $\sigma_Z,\sigma_Y$ are Pauli operators. We  choose $\alpha,\beta,\gamma,\delta$ randomly.  $R_Y,R_Z,XX$ are native gates of a trapped-ion quantum computer. } 
\label{fig:RQC}
\end{figure}

\subsection{Estimating the optimal number of copies}
\label{sec:nopt}
As REQUEST lifts the requirement for more and more qubits, it is especially important to determine the correct number of copies to use. 
We propose to estimate $M_{\textrm{opt}}$ by finding the optimal number of copies for  similar but classically simulable systems. To accomplish this we construct  a set of near-Clifford circuits that are similar to the state preparation circuit. (See Appendix~\ref{app:Cliffords} for details on how we define  similar circuits.) 

We will call the states prepared by these near-Clifford circuits $\{\ket{\Phi_i}\}$. If the observable of interest $X$ can be efficiently decomposed into a sum of Clifford operators, we are then able to efficiently classically compute (without noise) the exact expectation values of $X$ for these states. If we attempt to prepare the states $\{\ket{\Phi_i}\}$ on noisy hardware, we will end up instead preparing corresponding density matrices $\{\rho_i\}$. We expect and further back our claim up with numerical evidence in Section~\ref{sec:res} that the optimal number of copies for these near-Clifford circuits should be similar to the optimal number of copies for the circuit for which we would like to mitigate errors. Therefore, we approximate:
\begin{equation}
    M_{\textrm{opt}}\approx\min_M\Bigg\{\sum_i\Bigg|\frac{\Tr[X\rho_i'^M]}{\Tr[\rho_i'^M]}-\bra{\Phi_i}X\ket{\Phi_i}_\textrm{exact}\Bigg|\Bigg\}.
\end{equation}
Note that the traces $\Tr[X\rho_i'^M]$ and $\Tr[\rho_i'^M]$ are computed with the noisy quantum device while the expectation values $\bra{\Phi_i}X\ket{\Phi_i}_\textrm{exact}$ are computed classically.

\section{Resource Scaling}
\subsection{Overview}

Here we present scaling analysis that is applicable to both the VD and REQUEST methods.

Let $\mathcal{E}$ represent the size of the statistical errors one can tolerate in the estimation of $\Tr(X\rho^M)/\Tr(\rho^M)$. We now consider the scaling of the number of shots ($N_S$) required to reach this level of precision. In this case, the number of shots required with REQUEST scales similarly to the number of shots needed for VD. It has previously been shown that this scaling is polynomial in $1/\mathcal{E}$~\cite{koczor2020exponential}. This previous analysis did not, however, study the scaling of $N_S$ with the number of qubits $N$. We focus on that aspect here.

We consider the scaling of $N_S$ under the simplified assumptions that the controlled gates in Figure~\ref{fig:circuit_diagrams} are noiseless, the copies are identical, and the ancilla is unaffected by the noise. We first review the variance of the error mitigated quantity, and thus how many shots are required to reach a given precision. We next introduce a new upper bound on the number of shots required for generic noise models which don't violate the above assumptions. In particular, we show that the scaling can be bounded above by the scaling for  a global depolarizing noise model with the same error rate. Finally, we explicitly investigate the scaling for the bound on the number of shots required for a global depolarizing noise model. As global depolarizing noise would not introduce a noise floor, we use the results of~\cite{koczor2020exponential} to tie the scaling of the number of copies needed to get the accuracy (i.e., $\Tr(X\rho^M)/\Tr(\rho^M)-\ave{X}$) to be on the same order as the precision. We note, however, that the presence of the noise floor only changes the attainable accuracy, not the precision.

The bound for global depolarizing noise scales exponentially with $N$ for an error rate per circuit layer that is independent of the number of qubits and circuit with depth linear in $N$. If the error rate per layer instead increases with $N$ or one considers circuits with depth that scales faster, this bound becomes super-exponential. While this exponential or worse scaling does limit the regime in which either REQUEST or VD could be used, it does not rule out using these tools at the scale where quantum advantage is expected. Indeed our analysis finds that, for global depolarizing noise with reasonably small error rates, $N_S$ would remain feasible even for hundreds of qubits. This is larger than the scale needed to achieve quantum supremacy~\cite{google2019supremacy}.

\subsection{Variance Estimation}

We begin by computing the number of shots that need to be used to evaluate $\Tr(X\rho^M)/\Tr(\rho^M)$ to some fixed precision. Following the arguments of~\cite{koczor2020exponential} and taking a truncated Taylor series approximation in the large shot number limit, the variance of our estimator for $\Tr(X\rho^M)/\Tr(\rho^M)$ becomes:
\begin{equation}
\begin{aligned}
    \textrm{Var}\left(\frac{\Tr(X\rho^M)}{\Tr(\rho^M)}\right)\approx& \textrm{Var}\left(\textrm{prob}_0\right)\frac{4}{(2\textrm{prob}_0'-1)^2}\\
    &+\textrm{Var}\left(\textrm{prob}_0'\right)\frac{4(2\textrm{prob}_0-1)^2}{(2\textrm{prob}_0'-1)^4}.
\end{aligned}
\end{equation}
As the estimates for $\textrm{prob}_0$ and $\textrm{prob}_0'$ follow the binomial distribution we have that
\begin{equation}
    \textrm{Var}\left(\textrm{prob}_0\right)=\frac{\textrm{prob}_0(1-\textrm{prob}_0)}{N_s}
\end{equation}
and
\begin{equation}
    \textrm{Var}\left(\textrm{prob}_0'\right)=\frac{\textrm{prob}_0'(1-\textrm{prob}_0')}{N_s'}.
\end{equation}
Here $N_s$ and $N_s'$ are the number of shots used to estimate $\textrm{prob}_0$ and $\textrm{prob}_0'$ respectively. For simplicity, let us now set $N_s=N_s'$. 

To reach a target variance $\mathcal{E}^2$, we then require a number of shots
\begin{equation}
\begin{aligned}
    N_S\approx& \frac{4}{\mathcal{E}^2}\left( \frac{\textrm{prob}_0(1-\textrm{prob}_0)}{(2\textrm{prob}_0'-1)^2}\right.\\
    &\left.+\frac{(2\textrm{prob}_0-1)^2\textrm{prob}_0'(1-\textrm{prob}_0')}{(2\textrm{prob}_0'-1)^4}\right).
\end{aligned}
\end{equation}
Under the simplifying assumption of a noiseless ancilla, this result can be expressed in terms of $\Tr[X\rho^M]$ and $\Tr[\rho^M]$ as:
\begin{equation}\label{eq:gen_shots}
    N_S\approx\frac{\Tr[X\rho^M]^2+\Tr[\rho^M]^2-2\Tr[X\rho^M]^2\Tr[\rho^M]^2}{\Tr[\rho^M]^4\mathcal{E}^2}.
\end{equation}

\subsection{Bounding the Scaling for Measuring Pauli Products with General Noise Channels}\label{sec:Gen_scale}
The variance (and thus the number of shots required) will generally depend on the state and operator. To simplify the analysis we will assume that $X$ is a tensor product of Pauli operators. Under this assumption we note that $0\le \Tr[X\rho^M]^2\le\Tr[\rho^M]^2$. We now consider how to upper bound $N_S$ in Equation~\eqref{eq:gen_shots}, assuming that we can perform the necessary controlled swaps noiselessly. 

We will consider two cases. First, if $\Tr(\rho^M)\le\frac{1}{\sqrt{2}}$, $N_S$ is maximized when $\Tr[X\rho^M]^2$ is as large as possible. This gives an upper bound of:
\begin{equation}
    N_S\lesssim\frac{2}{\mathcal{E}^2}\left(\frac{1}{\Tr[\rho^M]^2}-1\right).
\end{equation}
However, if instead  $\Tr(\rho^M)>\frac{1}{\sqrt{2}}$, $N_S$ is maximized when $\Tr[X\rho^M]^2=0$:
\begin{equation}
    N_S\lesssim\frac{1}{\Tr[\rho^M]^2\mathcal{E}^2}.
\end{equation}
We therefore have the following bound $N_S$ for any physically valid value of $\Tr[\rho^M]$:
\begin{equation}\label{eq:gen_shots_bound}
\begin{aligned}
     N_S\lesssim&\frac{2}{\Tr[\rho^M]^2\mathcal{E}^2}\\
     \equiv &N_S^\textrm{max}
\end{aligned}
\end{equation}

Let us denote the density matrix we arrive at after a general channel $\rho_{\textrm{Gen}}$. We can bound $\Tr(\rho_{\textrm{Gen}}^M)$ from below in terms of the largest eigenvalue $p_1$:
\begin{equation}
\begin{aligned}
    \Tr[\rho_{\textrm{Gen}}^M]=&\sum_{i=1}^{2^N}p_i^M\\
    =&p_1^M+\sum_{i=2}^{2^N}p_i^M\\
    \ge& p_1^M+\frac{(1-p_1)^M}{(2^{N}-1)^{M-1}}.
\end{aligned}
\end{equation}
This inequality can be derived using the method of Lagrange multipliers with the constraint  $\sum_{i=1}^{2^N}p_i=1$. Noting that this lower bound is exact for the case of a global depolarizing channel, we can then write
\begin{equation}\label{eq:gen_bound_by_gd}
    \Tr(\rho_{\textrm{Gen}}^M)\ge\Tr(\rho_{\textrm{GD}}^M)
\end{equation}
where 
\begin{equation}
   \rho_{\textrm{GD}}=p_1\dya{\psi_1}+ \frac{1-p_1}{2^N-1}(\id-\dya{\psi_1})
\end{equation}
is the density matrix with the same error probability (i.e. the same value of ($1-p_1$)) that would arise about via a global depolarizing channel acting on the dominant eigenvector, $\ket{\psi_1}$.

Combining Equation~\eqref{eq:gen_bound_by_gd} with Equation~\eqref{eq:gen_shots_bound} then gives us
\begin{equation}\label{eq:shots_bound_by_gd}
    N_{S_{\textrm{Gen}}}^\textrm{max}\le N_{S_{\textrm{GD}}}^\textrm{max}
\end{equation}
where $N_{S_{\textrm{Gen}}}^\textrm{max}$ and $N_{S_{\textrm{GD}}}^\textrm{max}$ are the upper bound on the numbers of shots required for the general noise channel and the global depolarizing noise channel, respectively. We therefore have that the bound on the shot requirement scaling for global depolarizing noise also bounds the shot requirement scaling for any other model with the same error probability.


\subsection{Measuring Pauli Products with Global Depolarizing Channels}\label{sec:GD}

As we have the bound in Equation~\eqref{eq:shots_bound_by_gd}, we will now use global depolarizing noise to explore the resource scaling of REQUEST for growing system sizes. 

We consider a global depolarizing channel that acts after each layer of state preparation with a constant error rate $\delta$, and thus the total error rate will scale with the depth of the circuit. In this model, after $p$ layers of parallel gates intended to prepare the $N$ qubit pure state $\ket{\phi_p}$, we instead have a density matrix:
\begin{equation}\label{eq:rho_p}
    \begin{aligned}
        \rho_p=(1-\delta)^p\dya{\phi_p}+ \frac{1-(1-\delta)^p}{2^N}\id.
    \end{aligned}
\end{equation}
We note that if the error rate $\delta$ changes for different layers, the term $(1-\delta)$ above should be replaced by a geometric mean of that quantity over the different error rates. For this density matrix, we then have:
\begin{equation}\label{eq:Tr_rho_p}
\begin{aligned}
    \Tr[\rho_p^M]=&\sum_{k=0}^M {M \choose k} (1-\delta)^{p(M-k)}\left(\frac{1-(1-\delta)^p}{2^N}\right)^k\\
    &\,\,\,\,\,\,\,\,\,\,\,\cdot\Tr[(\dya{\phi_p})^{M-k}]\\
    =& \left(\frac{(1-\delta)^{p}(2^N-1)+1}{2^N}\right)^M\\ &+(2^N-1)\left(\frac{1-(1-\delta)^{p}}{2^N}\right)^M
\end{aligned}
\end{equation}
In the large $N$ limit we then have that this trace is exponentially suppressed with increasing circuit depth and number of copies:
\begin{equation}
    \lim_{N\to \infty}\Tr[\rho_p^M]=(1-\delta)^{pM}.
\end{equation}
We therefore find that $N_{S_{\textrm{GD}}}^\textrm{max}$ will asymptotically scale exponentially with increasing numbers of layers ($p$) and copies ($M$).

\begin{figure}[ht]
\includegraphics[width=.95\columnwidth]{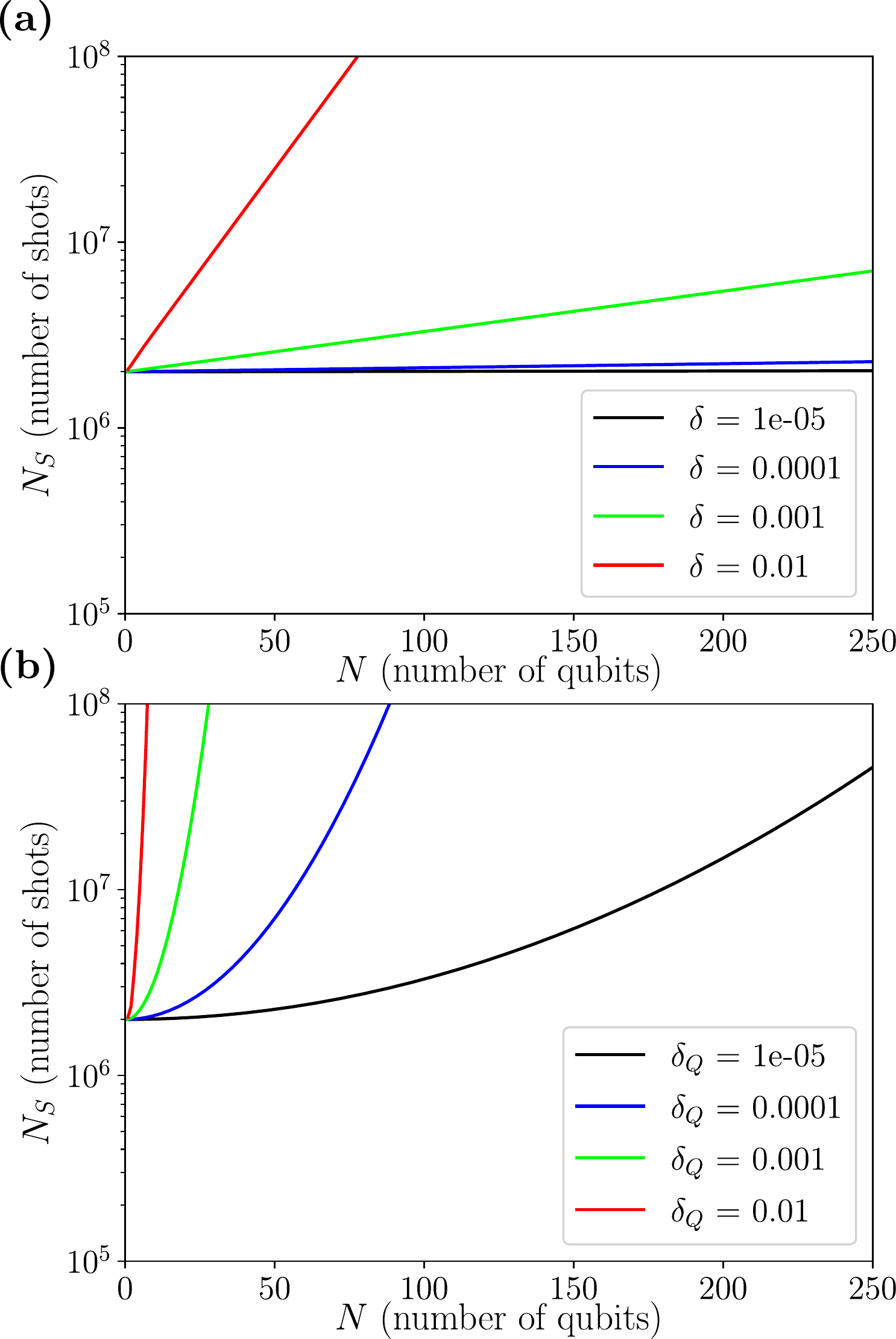}
\caption{\textbf{Upper bound on the shot cost scaling under different conditions.} In panels \textbf{(a)} and \textbf{(b)} we consider the number of shots required ($N_S$) under the assumption that the depth of the circuit grows linearly with the number of qubits $N$. In panel \textbf{(a)}  we assume that the error probability (denoted as $\delta$) per layer is independent of the number of qubits. In panel \textbf{(b)}  we instead model the error probability per layer as $\delta=(1-e^{-N\delta_Q})$. In all cases we ignore noise in the controlled swaps and target an error tolerance of $\mathcal{E}=0.001$.} 
\label{fig:scaling}
\end{figure}

We also need to consider what value of $M$ we should use for this scaling analysis. We note that even using a constant value of $M$ would still result in exponential scaling for $N_S$ so long as the circuit depth grows at least linearly. However, since we are working with global depolarizing noise, we can get the difference between the mitigated value and the exact value to be arbitrarily small by taking large enough values of $M$. With this in mind, we will use the minimal number of copies $M$ such that the approximation error is roughly $\mathcal{E}$. From~\cite{koczor2020exponential} we have that
\begin{equation}\label{eq:M_min}
    M=\left\lceil \frac{\textrm{ln}(2)+\textrm{ln}\left(\frac{1-p_1}{p_2}\right)-\textrm{ln}(\mathcal{E})}{\textrm{ln}\left(\frac{p_1}{p_2}\right)}\right\rceil
\end{equation}
where $p_1$ and $p_2$ are the first and second largest eigenvalues of $\rho_p$, respectively. Examining Equation~\eqref{eq:rho_p} we can see that in this case we have
\begin{equation}
    p_1=(1-\delta)^p+\frac{1-(1-\delta)^p}{2^N}
\end{equation}
and 
\begin{equation}
    p_2=\frac{1-(1-\delta)^p}{2^N}.
\end{equation}
Plugging these eigenvalues back into Equation~\eqref{eq:M_min} then gives
\begin{equation}\label{eq:M_p}
    M=\left\lceil \frac{\textrm{ln}(2^{N+1}-2)-\textrm{ln}(\mathcal{E})}{\textrm{ln}\left(2^N\frac{(1-\delta)^p}{1-(1-\delta)^p}+1\right)}\right\rceil.
\end{equation}

We can now use Equations~\eqref{eq:Tr_rho_p} and \eqref{eq:M_p} with Equation~\eqref{eq:gen_shots_bound} to estimate the upper bound on the number of shots needed for different qubit numbers $N$ for our global depolarizing model.


In Figure~\ref{fig:scaling}\textbf{(a)} we show how $N_{S_{\textrm{GD}}}^\textrm{max}$ would scale with $N$ for a case where $X$ is a Pauli product. For this demonstration, we take the circuit depth to be linear in the number of qubits, setting $p=N$, and assume a constant error rate with growing numbers of qubits. We observe that this situation leads to exponential growth in the shot cost with system size, but the cost grows slowly for  scales where a quantum advantage might be found ($N$ of order  $100$ qubits) and low enough error rate~\cite{neill2018blueprint}.

In Figure~\ref{fig:scaling}\textbf{(b)} we show similar results but no longer assume that the error rate can be held constant while we increase the number of qubits in a device. In order to model an increase that is approximately linear for small $N$ but saturates to 1 for large $N$, we set  $\delta=1-e^{-N\delta_Q}$. We consider this model as, empirically, larger quantum devices tend to be noisier than a smaller device with the same quality of qubits and gates. If this trend holds, we expect to find super-exponential scaling for $N_S$ similar to what is as shown in Figure~\ref{fig:scaling}\textbf{(b)}.

We note that this analysis has been carried out under the assumptions of noiseless controlled swaps, controlled-$X$ gate, and a noiseless ancilla. We did this in order to find the value of $N_{S_{\textrm{GD}}}^\textrm{max}$  that bounds $N_{S_{\textrm{Gen}}}^\textrm{max}$ under the same assumptions, as discussed in Section~\ref{sec:Gen_scale}. However, for the case of global depolarizing noise it is straightforward to drop these assumptions. Including noise in the controlled gates is simple as a global depolarizing channel commutes with them, meaning that this effectively increases the value of $p$ in the above analysis by an amount proportional to $N$. Dealing with noise on the ancilla is also simple. If a noiseless ancilla would be in the state $\rho_{\textrm{a}}$ right before measurement, the global depolarizing channel instead causes it to be in the state $\rho_{\textrm{a}}'$, given by:
\begin{equation}
    \rho_{\textrm{a}}'=(1-\delta)^p\rho_{\textrm{a}}+\frac{1-(1-\delta)^p}{2^N}\id.
\end{equation}
This then means that the measured value of $\Tr[\rho^M]$ is suppressed by a factor of $(1-\delta)^p$. Correspondingly, the upper bound on the number of shots required, $N_{S_{\textrm{GD}}}^\textrm{max}$, increases by a factor of $(1-\delta)^{-2p}$.

\subsection{Relation to Current Hardware}

Current trapped-ion quantum computers have error rates per gate $\delta_g=0.001-0.005$~\cite{cincio2020machine}. If we have order one gate per layer and assumed that idling noise in the layer is negligible, then we effectively have $\delta= \delta_g$, and one obtains that REQUEST can be used for systems as large as hundreds of qubits to obtain  $\mathcal{E}=0.001$ while increasing the cost only by a factor of order~$1$.
Here we compare the REQUEST shot cost to the cost of a single copy simulation with $\delta=0$. For circuits where the number of gates per layer roughly equals the number of qubits we expect that   $\delta\sim N\delta_g$ for sufficiently small  $\delta_g$. To model this case  we assume  the effective error rate per layer to be $\delta=1-e^{-N\delta_Q}$  with $\delta_Q = \delta_g$ and obtain that $\delta_g$ needs to be reduced by at most a factor of $10-50$ to obtain REQUEST mitigation for $N=100$ while increasing the cost only by a factor of order~$1$ and keeping $\mathcal{E}=0.001$.

\section{Numerical implementation}
\label{sec:res}

\begin{figure}[t]
\includegraphics[width=0.95\columnwidth]{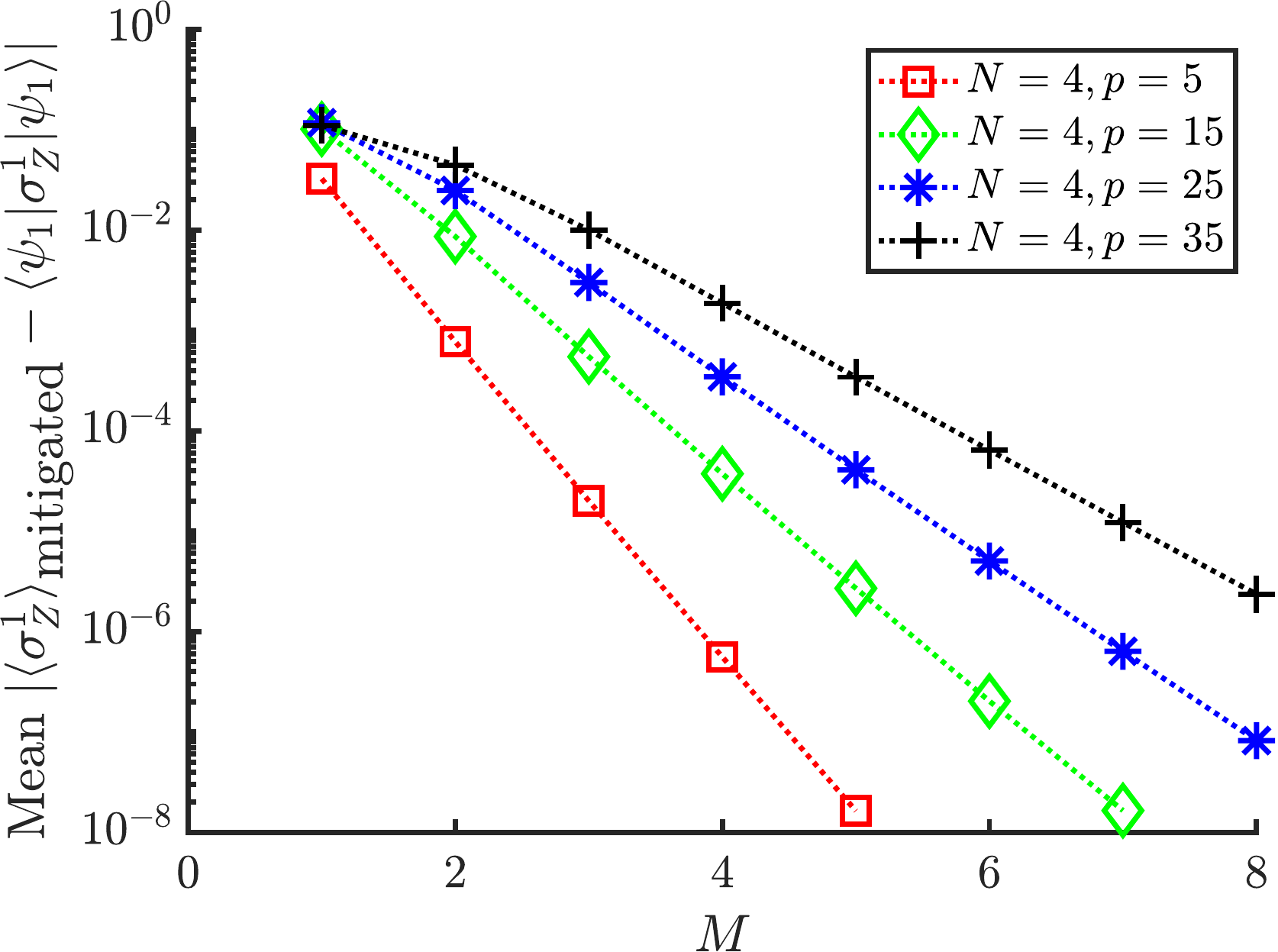}
\caption{\textbf{Exponential suppression of errors. }Mitigating $\avg{\sigma_Z^1}$ for RQC. Here, to clearly demonstrate the exponential suppression, we consider the noise acting only during the state preparation for the various copies. Furthermore, we plot the error with respect to $\avg{\sigma_Z^1}$ for $\ket{\psi_1}$.  The error is averaged over $44$ instances of RQC and plotted versus $M$. $M=1$ corresponds to noisy, single-copy results.} 
\label{fig:exp}
\end{figure}

We investigate the performance of both VD and REQUEST with a realistic noise model of trapped ion quantum computers \cite{trout2018simulating,cincio2020machine}. This architecture is favorable for REQUEST's  applications as the qubits have long decoherence times which limit the effects of idling noise during the derangement application.
Furthermore, such devices typically enable all-to-all qubit connectivity reducing circuit depth of the controlled derangements and observables. We assume all-to-all qubit connectivity in our implementation.

Our goal here is to introduce the REQUEST method and gain understanding of the effects of a  realistic noise model on the exponential error suppression methods.  Therefore, for simplicity  we choose to not combine the methods with other error mitigation methods. It seems probable that such a combination will further enhance the power of both VD and REQUEST.

\begin{figure}[t]
\includegraphics[width=0.95\columnwidth]{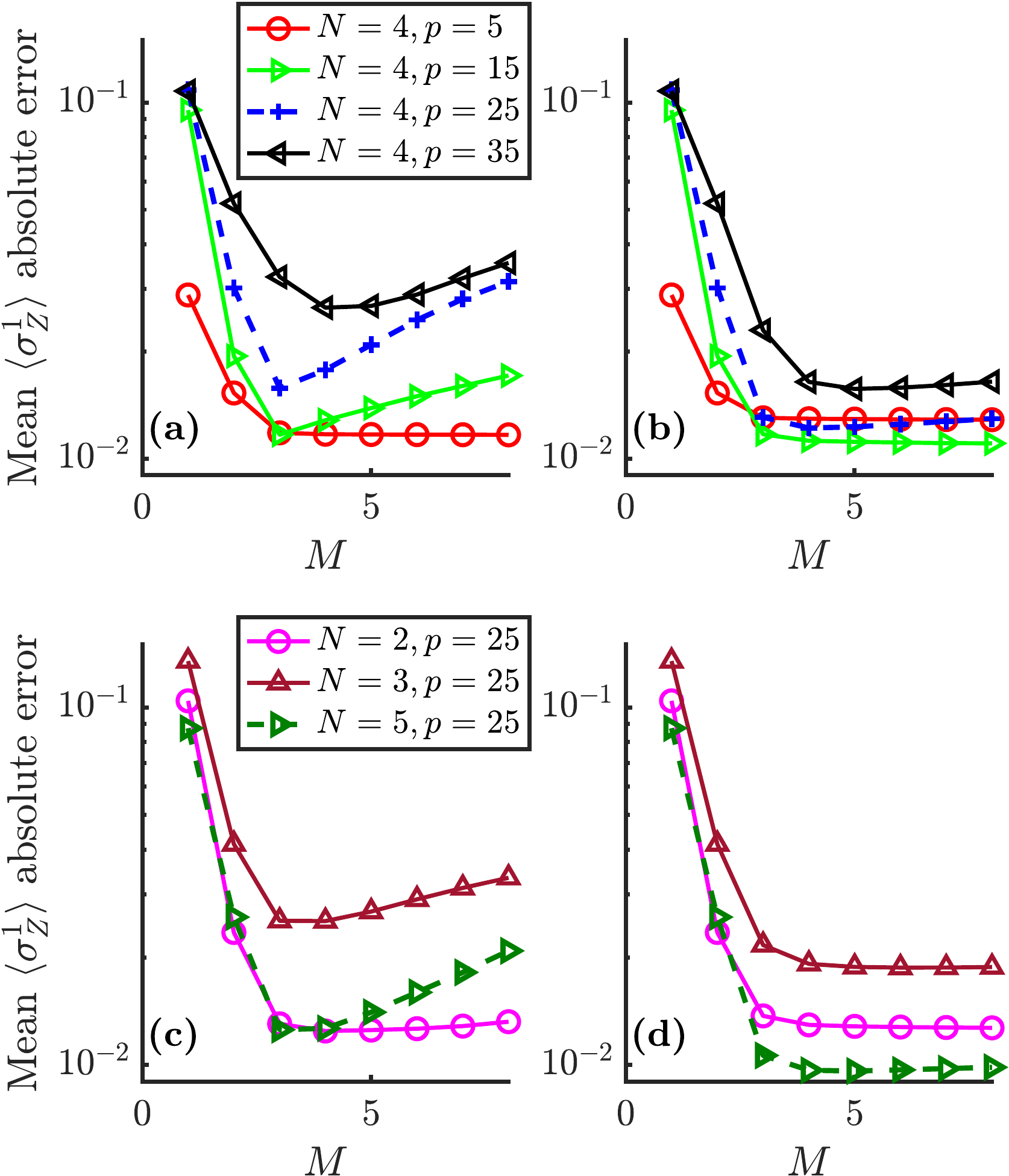}
\caption{\textbf{Error mitigation for random quantum circuits.} Error mitigation of $\avg{\sigma_Z^1}$ for the random quantum circuits with different $p$, $N$   and a realistic case when all gates are noisy. The error is averaged over $44$ instances of the random circuits.  In \textbf{(a)} and \textbf{(c)} we show the results obtained with the REQUEST method. The VD method is presented in panels \textbf{(b)} and \textbf{(d)}. Note that in our implementation both methods are equivalent for $M=2$. REQUEST allows us to obtain improvement over $M=2$ without extra qubits, providing the method's proof of principle.
Furthermore, for small enough $p$ the best results obtained with REQUEST and the original method have  similar quality.
}
\label{fig:res}
\end{figure}

To test the performance of the method we use random quantum circuits (RQC) obtained with a trapped-ion hardware efficient ansatz. The ansatz is built from layers of nearest-neighbor two-qubit $XX(\delta)$ gates that are parametrized by random angles $\delta$ and  decorated with general random single-qubit unitaries.  See Fig.~\ref{fig:RQC} for details. We test the method for a range of system sizes $N$ and numbers of ansatz layers $p$.

First, to clearly demonstrate the exponential suppression of errors, we consider a special  case when the noise acts only during the preparation of the various copies. Furthermore, we consider an error with respect to the leading eigenvector of the noisy state $\ket{\psi_1}$. We mitigate  $\avg{\sigma_Z^1}$ (the Pauli $Z$ operator on qubit 1) for RQC with $N=4$, $p=5,15,25,35$ averaging the error over 44 instances of RQC.  Indeed for  such a setup we find exponential suppression of errors, see Fig.~\ref{fig:exp}.   This is similar to the suppression observed in Ref.~\cite{koczor2020exponential}.

Next we consider REQUEST mitigation for a realistic case when all gates are noisy.
We mitigate $\avg{\sigma_Z^1}$ for RQC with $N=4$ and  $p=5,15,25,35$, and with $N=2,3,4,5$ and $p=25$.  We consider the mean absolute error of $\avg{\sigma_Z^1}$ (computed with respect to the exact expectation value $\langle\phi|\sigma_Z^1|\phi\rangle$) obtained by averaging over $44$ instances of RQC  plotted versus $M=1-8$. We gather the results  in Fig.~\ref{fig:res}\textbf{(a,c)}. 
In most cases we find a clearly visible $M_\textrm{opt}=3,4$ and  monotonic increase of the error for $M>M_\textrm{opt}$.  As $M_\textrm{opt}>2$, the results demonstrate advantage of the REQUEST method in  the case of  limited qubit counts.  

For benchmark purposes, we compare the results with the ones obtained by the  VD method, shown in Fig.~\ref{fig:res}\textbf{(b,d)}.  It is expected that VD will perform  better for large $M$ as the REQUEST circuit is much deeper for large $M$ for a device with full connectivity, such as the ion traps we simulate. (See Appendix~\ref{app:classical_VD} for details of the method used to classically simulate VD.) Nevertheless,  comparing the best results  obtained with both methods we see that they are similar (the errors differ by less than $35\%$), apart from the case of the deepest $p=35$ circuits for which VD outperforms REQUEST by a factor $1.7$.

\begin{figure}[t]
\includegraphics[width=0.95\columnwidth]{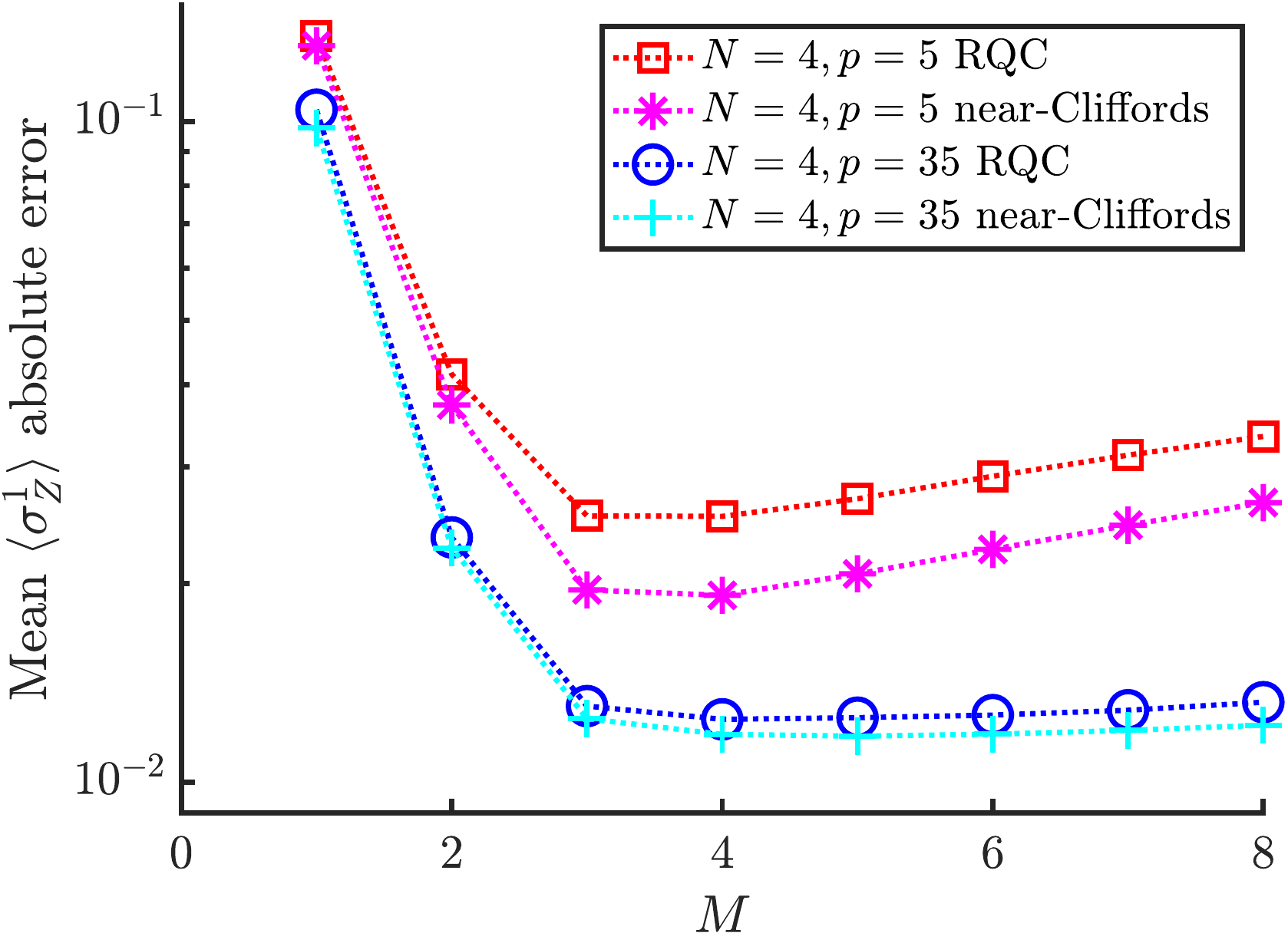}
\caption{ \textbf{Determining $M_{opt}$ with near-Clifford circuits.} A Comparison of the mean $\avg{\sigma_Z^1}$ absolute error  for REQUEST mitigation of  RQC  and near-Clifford circuits. RQC circuits are the ones for which results from Fig.~\ref{fig:res} were obtained  while the near-Clifford ones  are obtained by projecting  these RQC  as described in Section~\ref{sec:nopt} and Appendix~\ref{app:Cliffords}. Here the error is averaged over $44$ instances of the random circuits and $440$ instances of near-Clifford circuits.  The behavior is similar enough in both cases to use the near-Clifford circuits in order to  find a good approximation of $M_\textrm{opt}$ for the random circuits.  We show here the behavior for  $N$ and $p$ for which using the approximate $M_\textrm{opt}$ instead of the exact one results in the largest increase of the error (which is smaller than $1.5\%$).
 }
\label{fig:Cliffords}
\end{figure}

Furthermore, we test the method of finding $M_\textrm{opt}$ for REQUEST with the circuits used to obtain the results shown in Fig.~\ref{fig:res}.
We compare the behavior of the error averaged over RQC instances  as a function of $M$ in the cases of RQC and near-Clifford circuits,   see  Fig.~\ref{fig:Cliffords}.
We find both behaviors to be similar enough to successfully approximate the RQC's  $M_\textrm{opt}$ by its value for the near-Clifford circuits.
  
\begin{figure}[t!]
\includegraphics[width=0.95\columnwidth]{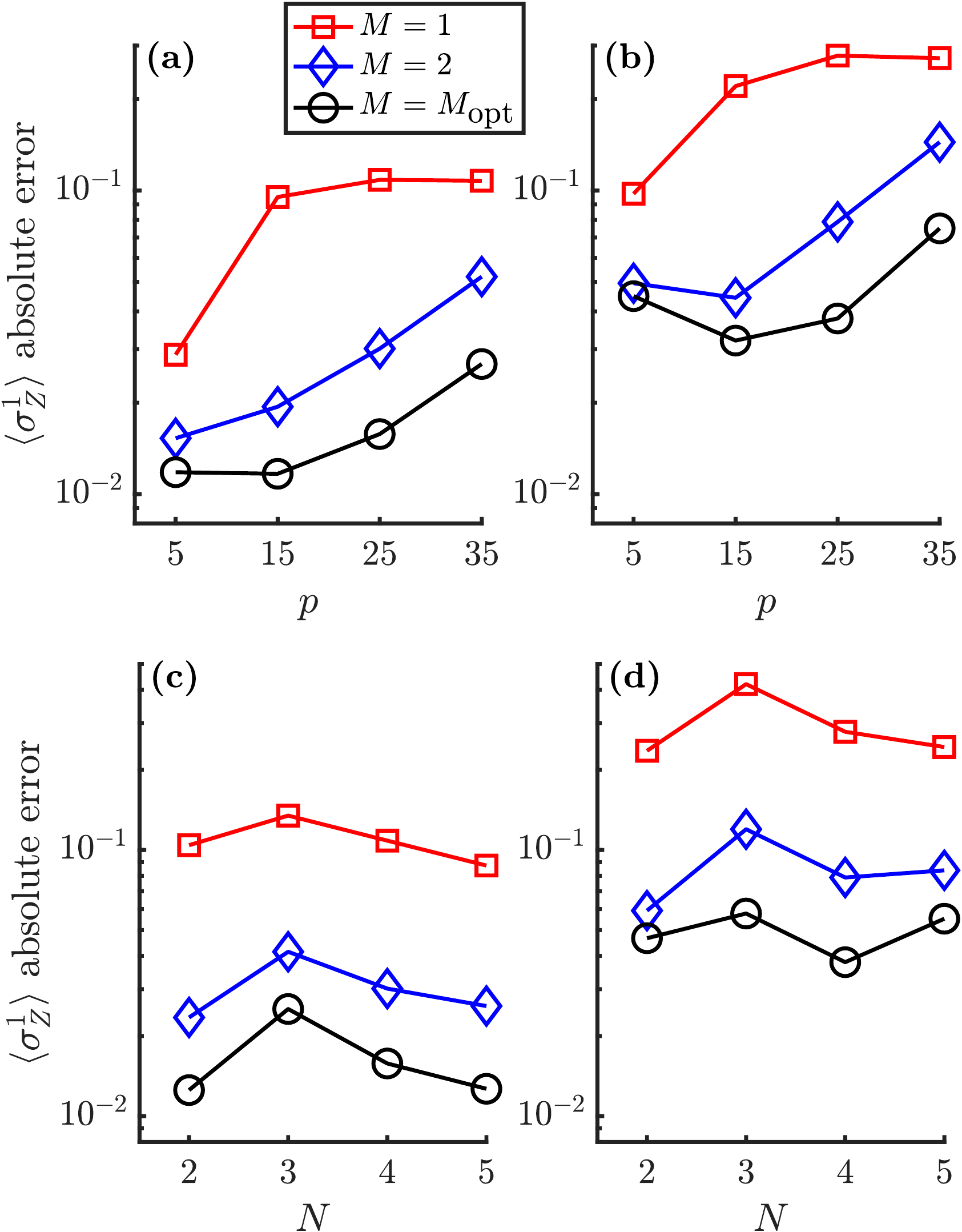}
\caption{\textbf{Summary of results obtained for RQC mitigation.} Scaling of $\avg{\sigma_Z^1}$ error for REQUEST RQC mitigation  from  Fig.~\ref{fig:res} and selected $M$ values. Panels \textbf{(a)}, \textbf{(c)} show  mean values of the  error, while panels \textbf{(b)}, \textbf{(d)} show the maximal values.  In \textbf{(a)}, \textbf{(b)} results for $N=4$, and  in \textbf{(c)}, \textbf{(d)} results for $p=25$.  $M_\textrm{opt}$ was estimated here with the near-Clifford circuits. The results demonstrate systematic improvement  obtained with the REQUEST method with respect to VD $M=2$ exponential mitigation and unmitigated results.       } 
\label{fig:scal}
\end{figure}

Finally, in Fig.~\ref{fig:scal} we gather the average and the maximal errors of $\avg{\sigma_Z^1}$ resulting from applying REQUEST to the circuits used in the comparison in Fig.~\ref{fig:res}. 
We compare the errors for  $M=1,\ldots,M_\textrm{opt}$.  In all considered cases $M=2$ results improve upon $M=1$ and $M=M_\textrm{opt}$ results improve upon $M=2$. With $M=M_\textrm{opt}$ we obtain up to a factor of $8$ improvement over $M=1$ ($N=4, p=15$)
and up to a factor of $2$ improvement over $M=2$  ($N=5, p=25$) for the average error. We observe that the quality of the mitigated results is decreasing with increasing $p$ while it is similar for all considered $N$.

\section{Conclusion}
The next major milestone for quantum computing is demonstrating an advantage over classical computing for some task that is of practical use. Achieving such a quantum advantage with NISQ devices will likely require algorithms that are qubit efficient. Additionally, as NISQ devices cannot support full error correction, quantum advantage will also require robust error mitigation strategies.

Building upon recent proposals to use multiple copies of a noisy state $\rho$ to distill the pure state of interest~\cite{koczor2020exponential,huggins2020virtual}, so-called Virtual Distillation (VD), we have presented a variation that achieves qubit efficiency by resetting and reusing qubits. Specifically, for $N$ qubit states, the total qubit requirement of our method, REQUEST, is only $2N+1$ for any number of copies while the previous approach required $MN+1$ qubits to use $M$ copies. 

As the number of copies used by REQUEST is not limited by the size of the physical device, we also address how to estimate the optimal number of copies, $M_\textrm{opt}$.  We propose to find $M_\textrm{opt}$ by using (classically simulable) near-Clifford circuits. Choosing these near-Clifford circuits to be similar to the circuit that prepares~$\rho$, one can compare results mitigated with different values of $M$ to exact quantities. 

We find that both VD and REQUEST require a number of shots that will scale exponentially in the number of qubits being considered. However, we find that with only modest increases in gate fidelities beyond what has been achieved with current hardware the exponential scaling does not rule out mitigating errors on hundreds of qubits. This result means that the exponential scaling does not prevent REQUEST from being helpful in the regime where practical quantum advantages are likely to be achieved.

While REQUEST achieves a reduction in qubit resources by increasing the overall depth of the quantum circuit, this trade-off can still be worthwhile.
Using a realistic trapped-ion noise model and random quantum circuits, we compare REQUEST and VD. For this test case, we find that the REQUEST method with $M_\textrm{opt}$ copies provides a clear advantage over VD with $M=2$ copies.

We note that when enough qubits are available on a fully connected device the qubit-hungry VD method tends to outperform REQUEST on $2N+1$ qubits when considering the same number of copies. However, while we have focused on the likely near-term case where few qubits are available, the active reset approach of REQUEST can be generalized further. If sufficient qubits and connectivity are available, REQUEST's active resets can be applied to more than one subsystem. Doing so would potentially increase the optimal number of copies $M_\textrm{opt}$ and with it the error suppression. REQUEST will therefore be relevant even when larger devices with good connectivity are available.

Beyond the NISQ regime, Ref.~\cite{huggins2020virtual} considers whether this kind of error suppression with surface codes might be useful. They note that if one needs to get the most physical qubits possible out of a device, low code distances would have to be considered. In such a situation, they argue that the exponential suppression technique with $M=2$ may offer a better improvement than using the extra qubits to increase the code distance. (This improvement vanishes for sufficiently large code distances.) As REQUEST can provide better results without requiring additional physical qubits, it would prove even more useful in that regime. We therefore expect that, even once fault tolerance is achieved, REQUEST will be useful for problems that are qubit limited. 

To further establish the method it will be important to benchmark its hardware implementation. Finally, the best way to combine these error suppression techniques with the established error mitigation strategies (such as zero noise extrapolation~\cite{temme2017error,kandala2018error,dumitrescu2018cloud,endo2018practical,otten2019recovering, giurgica2020digital,he2020zero}, Clifford data regression~\cite{czarnik2020error,lowe2020unified}, etc.) remains a question for future research.

\section{Acknowledgements}

We thank Yigit Subasi and Rolando Somma for insightful discussions. This work was supported by the Quantum Science Center (QSC), a National Quantum Information Science Research Center of the U.S. Department of Energy (DOE). Piotr C. and AA were also supported by the Laboratory Directed Research and Development (LDRD) program of Los Alamos National Laboratory (LANL) under project numbers 20190659PRD4 (Piotr C.) and 20210116DR (AA and Piotr. C). PJC also acknowledges initial support from the LANL ASC Beyond Moore's Law project. LC was also initially supported by the U.S. DOE, Office of Science, Office of Advanced Scientific Computing Research, under the Quantum Computing Application Teams~program.

\appendix

\section{Near-Clifford circuits construction}
\label{app:Cliffords}
To generate near-Clifford circuits used to determine $M_\textrm{opt}$ in Sec.~\ref{sec:res} we replace most  of the RQC gates by close to them Clifford gates. We use near-Clifford circuits with  $N_{\mathrm{nC}}=28$ non-Clifford gates. For such $N_{\mathrm{nC}}$  expectation values can be evaluated with current classical  near-Clifford simulators.   We adapt a Clifford substitution algorithm from \cite{czarnik2020error} to the case of trapped-ion devices. For the sake of simplicity we allow only near-Clifford  circuits with non-Clifford $R_Z(\alpha_j)$. 
Such choice is good enough to identify $M_\textrm{opt}$ in the case of RQC considered here. In general it might be beneficial to consider more general non-Clifford circuits.  We detail the algorithm below.

\begin{figure}[t]  \centering
\includegraphics[width=\columnwidth]{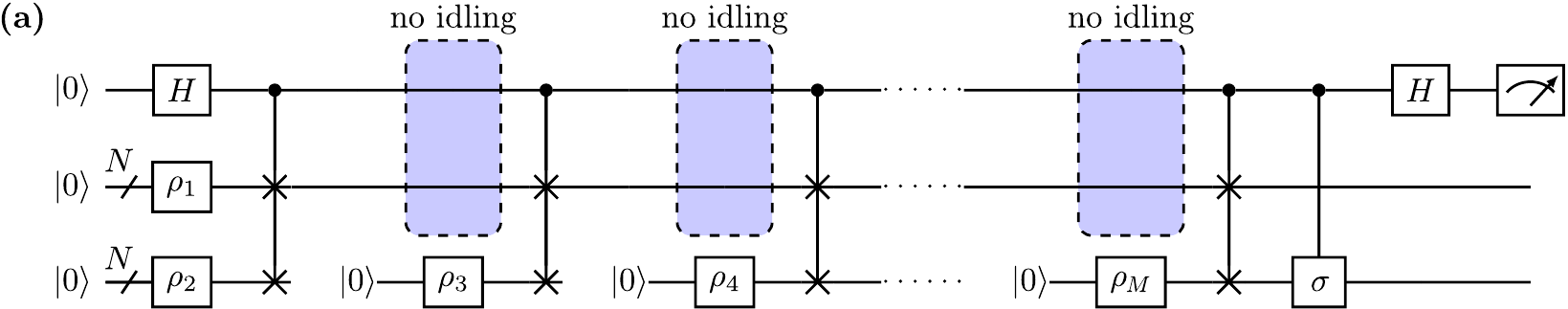} \\ \vspace{0.2cm}
\includegraphics[width=0.9\columnwidth]{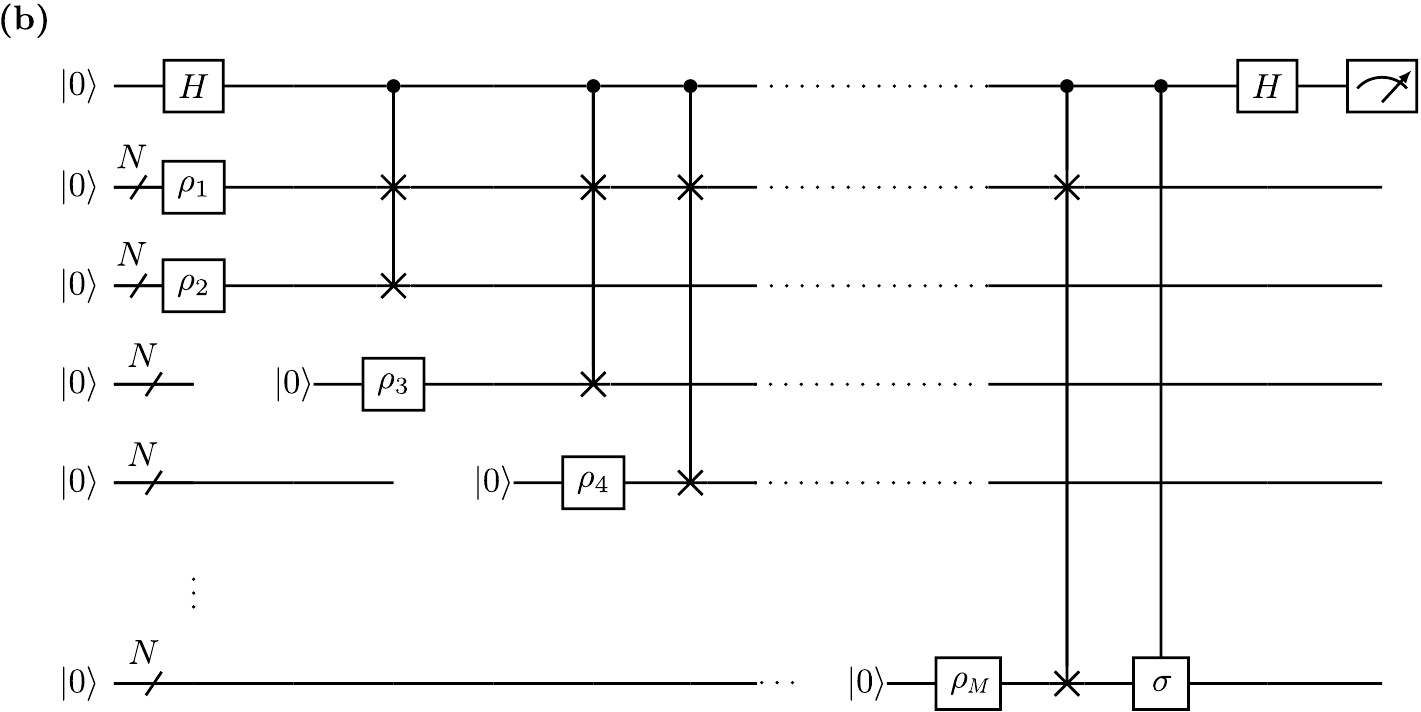} 
\caption{ \textbf{Classical simulation of VD mitigation.} In \textbf{(a)} a schematic representation of the qubit-efficient classical simulation of  VD. The circuit shown in (a) is simulated classically  with $2N+1$ qubits  while switching off an idling noise channel \cite{cincio2020machine} at the ancillary qubit and the first register of qubits (corresponding to  $\rho_1$) during preparation of  $\rho_2,\dots,\rho_M$. The switched off idling  gates correspond to the  blue areas in \textbf{(a)}.     The  computation shown in  \textbf{(a)} is equivalent to a noisy circuit  \textbf{(b)} implementing VD under assumptions (i) that there is no  cross-talk noise associated with preparation of $\rho_2,\dots,\rho_M$ and (ii) that depth of  circuits preparing $\rho_2,\dots,\rho_M$ is the same as depth of  circuits implementing CTRL-SWAPs. Assumption (ii) is necessary to ensure equivalent distribution of idling gates \cite{cincio2020machine} in both cases.  Assumption (i) is true for our noise model \cite{cincio2020machine}. Assumption (ii) is not true in general. In the case when (ii) is not true the equivalent circuit is similar to \textbf{(b)} although it can not be correctly visualized without considering structure of  native gates in the circuits implementing  CTRL-SWAPs and  state preparation.  Finally, note that we apply the controlled-$\sigma$ operation to $\rho_M$. While in the idealized case of noise acting only during state preparation and  $\rho_1=\rho_2=\dots=\rho_M$  all choices of the measured copy are equivalent that's no longer true for realistic noise. We find that   for our implementation   choosing $\rho_M$  minimizes the error.    } 
\label{fig:classical}
\end{figure}

\begin{enumerate}
\item 
Decompose a random quantum circuit to the native gates of a trapped-ion computer. In the case of our RQC implementation  the decomposition is obtained with  $R_Z(\alpha), R_Y(\beta), XX(\delta)$ as described in Fig.~\ref{fig:RQC}.
\item 
Identify all non-Clifford gates and their angles $\alpha_j, \beta_j, \delta_j$. 
\item
For each non-Clifford gate  generate weights $w$ determining the probability of its replacement by Clifford gates.
As $Z$ rotations by an angle of $k\pi/2,\, k=0,1,2,3$ are in the Clifford group,  we consider making the substitution  $R_Z(\alpha_j) \to R_Z(k\pi/2),$ with weights
\begin{displaymath}
w_{jk} =  e^{-d^2/\sigma^2},\ d = \frac{||e^{i\alpha_j/2} R_Z(\alpha_j) - e^{ik\pi/4}R_Z(k\pi/2)||}{||R_Z(\alpha_j)||},
\end{displaymath}
for the different $k$ values. Here $||.||$ is a Frobenius matrix norm and $\sigma$ is an adjustable parameter. 
Similarly, we consider replacing the angles in $Y$ rotations to bring the rotation into the Clifford group:  $R_Y(\beta_j) \to R_Y(k\pi/2),\, k=0,1,2,3$  with weights given by
\begin{displaymath}
w_{jk} = e^{-d^2/\sigma^2},\ d = \frac{||e^{i\beta_j/2} R_Y(\beta_j) - e^{ik\pi/4}R_Y(k\pi/2)||}{||R_Y(\beta_j)||}.
\end{displaymath}
Finally, we also consider replacing  $XX(\delta_j)$ gates with the Clifford rotations  $XX(k\pi/4),\, k=0,1,2,3$ with the weights
\begin{displaymath}
w_{jk} =  e^{-d^2/\sigma^2},\ d = \frac{||e^{i\delta_l} XX(\delta_j) - e^{ik\pi/4}XX(k\pi/4)||}{||XX(\delta_j)||}.
\end{displaymath}
We choose $\sigma=0.5$.
\item
We replace each $R_Y(\beta_j)$ ($XX(\delta_j)$) by Clifford $R_Y(k\pi/2)$ ($XX(k\pi/4)$) with probability 
\begin{displaymath}
p_{jk} = \frac{w_{jk}}{\sum_k w_{jk}}.
\end{displaymath}
\item\label{replace_step}
We replace one of $R_Z(\alpha_j)$ by $R_Z(k\pi/2)$ with probability 
\begin{displaymath}
p_{jk} = \frac{w_{jk}}{\sum_{jk} w_{jk}}.
\end{displaymath}
\item
If after the replacement the number of non-Clifford  $R_Z(\alpha_j)$ is larger than $N_{\mathrm{nC}}$,  we  repeat step~\ref{replace_step} until only $N_{\mathrm{nC}}$ non-Clifford $R_Z(\alpha_j)$ remains.
\end{enumerate}
To determine $M_{opt}$ we generate $4-10$ near-Clifford circuits per each random quantum circuit.

\section{Implementation details}
\label{app:implementation}

\subsection{Quantum gates decomposition  to native trapped-ion gates. }

To implement the CTRL-SWAP gate we decompose it to two CNOTs and a Toffoli gate as described in \cite{smolin1996five}. We decompose Toffoli gates, CNOTs and CTRL-Z gates to the native gates using  a decomposition from  \cite{maslov2017basic} and assuming full connectivity of the device.

\subsection{Classical simulation of the VD method}
\label{app:classical_VD}
Noisy classical simulation of VD is challenging for large $M$ as the number of the required qubits  grows linearly with $M$. Nevertheless in absence of cross-talk noise (as in the case of our noise model)  an equivalent classical  simulation can be performed with $2N+1$ qubits, see Fig.~\ref{fig:classical}. We use a method from  Fig.~\ref{fig:classical} to simulate VD.

\bibliography{quantum.bib}
\end{document}